\documentclass[aps,pra,amssymb,amsmath,eqsecnum,nofootinbib]{revtex4}

\newtheorem{definition}{Definition}[section]
\newtheorem{theorem}{Theorem}[section]

\newtheorem{corollary}{Corollary}[section]
\newtheorem{conjecture}{Conjecture}[section]
\newtheorem{remark}{Remark}[section]

\newenvironment{hypothesis}{HP: \begin{center}} {\end{center}}
\newenvironment{thesis}{TH: \begin{center}} {\end{center}}
\newenvironment{proof}{\begin{center}PROOF: \end{center}} {$ \blacksquare $}
\newtheorem{example}{Example}[section]
\begin{document}
\title{Classification of the automorphisms of the
    noncommutative torus among the (chaotic and non-chaotic) shallow ones
    and the non-chaotic complex ones}
\author{Gavriel Segre}
\begin{abstract}
Adopting the measure of quantum complexity, the quantum logical depth, previously introduced by the author
the  automorphisms of the noncommutative torus  are classified among the (chaotic and non-chaotic) shallow ones and the non-chaotic complex ones.
\end{abstract}
\maketitle
\newpage
\tableofcontents
\newpage
\section{Acknowledgements}
I would like to thank Giovanni Jona-Lasinio and Vittorio de Alfaro for many useful suggestions.

Of course nobody among them has any responsibility a to any (eventual) mistake contained in these pages.
\newpage
\section{Introduction} \label{sec:Introduction}
No term in Science is used in a plethora of completely different meanings as the term \emph{complexity}.

Indeed the locution \emph{Science of Complex Systems} is very often used as a label intentionally left undefined in order to comprehend any kind of research program.

A mathematical precise measure of  complexity, called \emph{logical depth}, was introduced by Charles Bennett \cite{Bennett-88} and generalized by myself to quantum systems \cite{Segre-04b}.

A great intellectual achievement of such a notion consists in the fact that it clearly distinguishes between \emph{complexity} and \emph{randomess}: a random system is no complex.

Since, by \emph{Brudno theorem} \cite{Brudno-78}, \cite{Brudno-83}, \cite{Alekseev-Yakobson-1981},  the dynamical entropy of a classical dynamical system is equal to the \emph{algorithmic information} \cite{Calude-02} (that despite being usually  also called  \emph{algorithmic complexity} or \emph{Kolmogorov complexity} \cite{Li-Vitanyi-97} in a misleading way is  not  a measure of complexity) of almost every trajectory simbolically codified  and since by definition a dynamical system is chaotic if it has positive dynamical entropy it follows that a complex dynamical system is not chaotic.

 The generalization of  Brudno Theorem to the quantum case  recentely performed by Fabio  Benatti, Tyll Kruger, Markus Muller, Rainer Siegmund-Schultze and Arleta Szkola   \cite{Benatti-Kruger-Muller-Siegmund-Schultze-Szkola-06}, \cite{Benatti-09} implies that also at the quantum level \emph{quantum chaos} implies no \emph{quantum complexity}.

This is a rather astonishing fact since \emph{quantum chaoticity} and \emph{quantum complexity} are very often erroneously used as synonymous.

In this paper we show this fact by considering the example consisting in the automorphisms of the noncommutative torus and classifying them among the (chaotic and non-chaotic) shallow ones and the non-chaotic complex ones.
\newpage
\section{The definition of the physical complexity of strings and sequences of
cbits} \label{sec:The definition of the physical complexity of
strings and sequences of cbits}

I will follow from here and beyond the notation of my Phd-thesis
\cite{Segre-02}; consequentially, given the binary alphabet $
\Sigma \, := \, \{ 0 \, , \, 1 \} $, I will denote by $
\Sigma^{\star} $ the set of all the strings on $ \Sigma $ (i.e.
the set of all the strings of cbits), by  $ \Sigma^{\infty} $ the
set of all the sequences on  $ \Sigma $ (i.e. the set of all the
sequences of cbits) and  by $ CHAITIN-RANDOM ( \Sigma^{\infty} )$
its subset consisting of all the Martin-L\"{o}f-Solovay-Chaitin
random sequences of cbits.

I will furthermore denote strings by an upper arrow and sequences
by an upper bar, so that I will talk about the string $ \vec{x}
\in  \Sigma^{\star} $ and the sequence $ \bar{x} \in
\Sigma^{\infty}$; $ | \vec{x} | $ will denote the length of the
string $ \vec{x} $, $ x_{n} $ will denote the $ n^{th}$-digit of
the string  $ \vec{x} $  or of the sequence $ \bar{x}$  while $
\vec{x}_{n} $ will denote its prefix of length n.

I will, finally, denote by  $ <_{l}$ the lexicographical-ordering relation
over $ \Sigma^{\star} $ and by  string(n)  the $ n^{th} $ string in
such an ordering.

Fixed once and for all a universal Chaitin computer U, let us
recall the following basic notions:

given a string $ \vec{x} \in \Sigma^{\star} $ and a natural number
$ n \in { \mathbb{N}} $:
\begin{definition} \label{def:canonical program of a string}
\end{definition}
\emph{canonical program of $ \vec{x}  $}:
\begin{equation}
  \vec{x}^{\star} \; := \; \min_{<_{l}} \{
  \vec{y} \in \Sigma^{\star} \, : \, U( \vec{y} ) =  \vec{x} \}
\end{equation}
\begin{definition}
\end{definition}
\emph{$ \vec{x}$ is n-compressible}:
\begin{equation}
    | \vec{x}^{\star} | \; \leq \; | \vec{x} | \, - \, n
\end{equation}
\begin{definition}
\end{definition}
\emph{$ \vec{x}$ is n-incompressible}:
\begin{equation}
    | \vec{x}^{\star} | \; > \; | \vec{x} | \, - \, n
\end{equation}
\begin{definition}
\end{definition}
\emph{halting time of the computation with input $ \vec{x} $}:
\begin{equation}
    T( \vec{x} ) \; := \; \left\{%
\begin{array}{ll}
    \text{number of computational steps after which U halts on input $
\vec{x} $ }, & \hbox{if $ U(\vec{x}) = \downarrow$} \\
    + \infty, & \hbox{otherwise.} \\
\end{array}%
\right.
\end{equation}
We have at last all the ingredients required to introduce the
notion of \emph{logical depth} as to strings.

Given a string $ \vec{x} \in \Sigma^{\star} $ and two natural
number $ s , t \in {\mathbb{N}} $:
\begin{definition} \label{def:logical depth of a string of cbits}
\end{definition}
\emph{logical depth of  $ \vec{x} $ at significance level s}:
\begin{equation}
    D_{s}(  \vec{x} ) \; := \; \min \{ T( \vec{y} ) \, : \, U( \vec{y}
    ) \, = \, \vec{x} \, , \, \vec{y} \text{ s-incompressible } \}
\end{equation}
\begin{definition}
\end{definition}
\emph{$ \vec{x} $ is t-deep at significance level s}:
\begin{equation}
   D_{s}(  \vec{x} ) \; > \; t
\end{equation}
\begin{definition}
\end{definition}
\emph{$ \vec{x} $ is t-shallow at significance level s}:
\begin{equation}
   D_{s}(  \vec{x} ) \; \leq  \; t
\end{equation}

I will denote the set of all the  t-deep strings as $ t-DEEP(
\Sigma^{\star} ) $ and the set of all the t-shallow strings as $
t-SHALLOW( \Sigma^{\star} ) $ .

Exactly  as it is impossible to give a sharp distinction among
\emph{Chaitin-random} and \emph{regular} strings while it is
possible to give a sharp distinction among
\emph{Martin-L\"{o}f-Solovay-Chaitin-random} and \emph{regular}
sequences, it is impossible to give a sharp distinction among
\emph{deep} and \emph{shallow} strings  while it is possible to
give a sharp distinction among \emph{deep} and \emph{shallow}
sequences.

Given a sequence $ \bar{x} \in \Sigma^{\infty} $:
\begin{definition} \label{def:strongly-deep sequence of cbits}
\end{definition}
\emph{$ \bar{x} $ is strongly deep}:
\begin{equation}
    card \{ n \in {\mathbb{N}} \, : \, D_{s}( \vec{x}(n) ) > f(n)
    \} \; < \; \aleph_{0} \; \; \forall s \in {\mathbb{N}} \, , \,
    \forall f \in REC-MAP ( {\mathbb{N}} \, , \, {\mathbb{N}} )
\end{equation}
where, following once more the notation adopted in
\cite{Segre-02}, $ REC-MAP ( {\mathbb{N}} \, , \, {\mathbb{N}} ) $
denotes the set of all the (total) recursive functions over $
{\mathbb{N}}$.

To introduce a weaker notion of depth it is necessary to fix the
notation as to reducibilities and degrees:

denoted the \emph{Turing reducibility} by $ \leq_{T} $ and the
polynomial time Turing reducibility by  $ \leq_{T}^{P} $
\cite{Odifreddi-89} let us recall that there is an intermediate
constrained-reducibility among them: the one, called
\emph{recursive time bound reducibility}, in  which the
halting-time is constrained to be not necessarily a polynomial but
a generic recursive function; since \emph{recursive time bound
reducibility} may be proved to be equivalent to  \emph{truth-table
reducibility} (I demand to \cite{Odifreddi-99a},\cite{Calude-02}
for its definition and for the proof of the equivalence) I will
denote it by $ \leq_{tt} $.

A celebrated theorem proved  by Peter Gacs in 1986 \cite{Gacs-86}
states that every sequence is computable by a Martin
L\"{o}f-Solovay-Chaitin-random sequence:
\begin{theorem}
\end{theorem}
\emph{Gacs' Theorem}:
\begin{equation}
    \bar{x} \;  \leq_{T} \; \bar{y} \; \; \forall \bar{x} \in
    \Sigma^{\infty} \, , \, \forall \bar{y} \in
    CHAITIN-RANDOM (\Sigma^{\infty} )
\end{equation}
This is no more true, anyway, if one adds the constraint of
recursive time bound, leading to the following:
\begin{definition} \label{def:weakly-deep sequence of cbits}
\end{definition}
\emph{$ \bar{x} $ is weakly deep}:
\begin{equation}
    \exists \bar{y} \in CHAITIN-RANDOM (\Sigma^{\infty} ) \, : \neg ( \bar{x} \;
  \leq_{tt} \; \bar{y} )
\end{equation}
I will denote the set of all the strongly-deep binary sequences by
$ STRONGLY-DEEP(\Sigma^{\infty} ) $ and the set of all the
weakly-deep binary sequences as $ WEAKLY-DEEP(\Sigma^{\infty}) $.

Shallowness is then once more defined as the opposite of depth:
\begin{definition}
\end{definition}
\emph{strongly-shallow sequences of cbits}:
\begin{equation}
  STRONGLY-SHALLOW(\Sigma^{\infty} ) \; := \; \Sigma^{\infty} \, -
  \, (STRONGLY-DEEP(\Sigma^{\infty} ))
\end{equation}
\begin{definition}
\end{definition}
\emph{weakly-shallow sequences of cbits}:
\begin{equation}
  WEAKLY-SHALLOW(\Sigma^{\infty} ) \; := \; \Sigma^{\infty} \, -
  \, (WEAKLY-DEEP(\Sigma^{\infty} ))
\end{equation}

Weakly-shallow  sequences of cbits may also be characterized in
the following useful way \cite{Bennett-88}:
\begin{theorem} \label{th:alternative characterization of weakly-shallow
sequences of cbits}
\end{theorem}
\emph{Alternative characterization of weakly-shallow sequences of
cbits}:
\begin{equation}
  \bar{x} \in WEAKLY-SHALLOW(\Sigma^{\infty} ) \; \Leftrightarrow
  \; \exists \mu \, recursive \; : \; \bar{x} \in \mu-RANDOM( \Sigma^{
\infty
    })
\end{equation}
where, following once more the notation  of \cite{Segre-02},  $
\mu-RANDOM( \Sigma^{ \infty} ) $ denotes the set of all the
Martin-L\"{o}f random sequences w.r.t. the measure $ \mu $.

As to sequences of cbits, the considerations made in
section\ref{sec:Introduction} may be
thoroughly formalized through the following:
\begin{theorem} \label{th:weak-shallowness of random sequences of cbits}
\end{theorem}
\emph{Weak-shallowness of Martin L\"{o}f - Solovay - Chaitin
random sequences}:
\begin{equation}
    CHAITIN-RANDOM( \Sigma^{ \infty} ) \, \cap \,
    WEAKLY-DEEP(\Sigma^{\infty}) \; = \; \emptyset
\end{equation}
\begin{proof}
Since the Lebesgue measure $ \mu_{Lebesgue} $ is recursive and by
definition:
\begin{equation}
     CHAITIN-RANDOM( \Sigma^{ \infty} ) \; = \; \mu_{Lebesgue} - RANDOM(
\Sigma^{ \infty} )
\end{equation}
the thesis immediately follows by the theorem \ref{th:alternative
characterization of weakly-shallow sequences of cbits}.
\end{proof}
\newpage
\section{The definition of the physical complexity of classical dynamical
systems} \label{sec:The physical complexity of classical dynamical
systems}

Since much of the fashion about complexity is based on a
spread confusion among different notions, starting from the basic
difference among \emph{plain Kolmogorov complexity} K and
\emph{algorithmic information} I, much care has to be taken.

Let us start from the following notions by Brudno:

\begin{definition} \label{def:Brudno algorithmic information of a sequence}
\end{definition}
\emph{Brudno algorithmic entropy of $ \bar{x} \in \Sigma^{\infty}
$}:
\begin{equation}
  B( \bar{x} ) \; := \; \lim_{n \rightarrow \infty} \frac{K( \vec{x}(n))
}{n}
\end{equation}

\smallskip
At this point one could think that considering the asympotic rate
of \emph{algorithmic information} instead of \emph{plain
Kolmogorov complexity} would result in a different definition of
the algorithmic entropy of a sequence.

That this is not the case is the content of the following:
\begin{theorem} \label{th:plain-prefix insensitivity of Brudno algorithmic
entropy}
\end{theorem}
\begin{equation}
   B( \bar{x} ) \; = \; \lim_{n \rightarrow \infty} \frac{I( \vec{x}(n))}{n}
\end{equation}
\begin{proof}
It immediately follows by the fact that \cite{Staiger-99}:
\begin{equation}
  | I( \vec{x}(n) ) \, - \, K( \vec{x}(n) )| \; \leq \; o(n)
\end{equation}
\end{proof}

\begin{definition} \label{def:Brudno random sequence}
\end{definition}
\emph{$ \bar{x} \in \Sigma^{\infty} $ is Brudno-random}:
\begin{equation}
  B( \bar{x} ) \; > \; 0
\end{equation}
I will denote the set of all the Brudno random binary sequences by
$ BRUDNO( \Sigma^{\infty})  $.

One  great source of confusion in a part of the literature arises
from the ignorance of the following basic result proved by Brudno
himself \cite{Brudno-78}:
\begin{theorem} \label{th:Brudno randomness is weaker than Chaitin
randomness}
\end{theorem}
\emph{Brudno randomness is weaker than Chaitin randomness}:
\begin{equation}
  BRUDNO-RANDOM( \Sigma^{\infty}) \; \supset \;  CHAITIN
-RANDOM(\Sigma^{\infty})
\end{equation}
as we will see in the sequel of this section.

Following the analysis performed in \cite{Segre-02} (to which I
demand for further details) I will recall here some basic notion
of Classical Ergodic Theory:

given a classical probability space $ ( X \, , \, \mu ) $:
\begin{definition}  \label{def:endomorphism of a classical probability
space}
\end{definition}
\emph{endomorphism of  $ ( X \, , \, \mu ) $}:

$T \, : \, HALTING(\mu) \rightarrow HALTING(\mu) $ surjective :
\begin{equation}
  \mu ( A ) \; = \;   \mu ( T^{-1} A ) \; \; \forall A \in HALTING(\mu)
\end{equation}
where $ HALTING(\mu) $ is the halting-set of the measure $ \mu $,
namely the $ \sigma$-algebra of subsets of X on which $ \mu $ is
defined.

\newpage
\begin{definition} \label{def:classical dynamical system}
\end{definition}
\emph{classical dynamical system}:

a triple $ ( X \, , \, \mu \, , \, T ) $ such that:
\begin{itemize}
  \item  $ ( X \, , \, \mu ) $ is a classical probability space
  \item  $T \, : \, HALTING(\mu) \rightarrow HALTING(\mu) $ is an
  endomorphism of  $ ( X \, , \, \mu ) $
\end{itemize}
Given a classical dynamical system  $ ( X \, , \, \mu \, , \, T )
$:

\begin{definition} \label{def:ergodic classical dynamical system}
\end{definition}
\emph{$ ( X \, , \, \mu \, , \, T ) $ is ergodic}:
\begin{equation}
lim_{n \rightarrow \infty} \frac{1}{n} \sum_{k=0}^{n-1} \, \mu ( A
\cap T^{k}(B)) \; = \; \mu(A) \,  \mu(B) \; \forall \, A,B \in
HALTING( \mu )
\end{equation}

\begin{definition}
\end{definition}
\emph{n-letters alphabet}:
\begin{equation}
  \Sigma_{n} \; := \; \{ 0 , \cdots , n - 1 \}
\end{equation}
Clearly:
\begin{equation}
  \Sigma_{2} \; = \; \Sigma
\end{equation}

Given a classical probability space $ ( X \, , \mu ) $:
\begin{definition} \label{def:partition of a classical probability space}
\end{definition}
\emph{finite measurable partition of $ ( X \, , \, \mu ) $}:
\begin{equation}
\begin{split}
  A \, &  = \; \{ \, A_{0} \, , \, \cdots \, A_{n-1} \} \; n \in
{\mathbb{N}} \, : \\
  A_{i} & \in  HALTING(\mu) \; \; i \, = \, 0 \, , \, \cdots  \, , \, n-1 \\
  A_{i} & \, \cap \, A_{j} \, = \, \emptyset \; \; \forall \, i \, \neq \,
  j \\
  \mu  &  ( \, X  \,- \, \cup_{i=0}^{n-1} A_{i} \, ) \; = \; 0
\end{split}
\end{equation}
I will denote the set of all the finite measurable partitions of $
( X \, , \, \mu ) $ by $ \mathcal{P} ( \, X \, , \, \mu \, ) $.

Given two partitions $ A = \{ A_{i} \}_{i=0}^{n-1} \, , \,  B  =
\{ B_{j} \}_{j=0}^{m-1} \; \in \; \mathcal{P} ( X , \mu ) $:
\begin{definition}
\end{definition}
\emph{A is a coarse-graining of B  $ (A \preceq B) $}:

every atom of A is the union of atoms by B

\smallskip

\begin{definition}
\end{definition}
\emph{coarsest refinement of  $ A = \{ A_{i} \}_{i=0}^{n-1} $ and
$ B = \{ B_{j} \}_{j=0}^{m-1} \in {\mathcal{P}}( \; X \, , \mu \;
) $}:
\begin{equation}
  \begin{split}
      A \, & \vee \, B \; \in {\mathcal{P}}( X  , \mu )  \\
      A \, & \vee \, B \; := \; \{ \, A_{i} \, \cap \, B_{j} \, \;  i =0 ,
\cdots , n-1 \; j = 0 , \cdots , m-1  \}
  \end{split}
\end{equation}

Clearly $ \mathcal{P} ( X , \mu ) $ is closed both under coarsest
refinements and under endomorphisms of $ ( X , \mu ) $.

Let us observe that, beside its abstract, mathematical
formalization, the definition\ref{def:partition of a classical
probability space} has a precise operational meaning.

Given the classical probability space $  ( X , \mu ) $ let us
suppose to make an experiment on the probabilistic universe it
describes using an instrument whose distinguishing power is
limited in that it is not able to distinguish events belonging to
the same atom of a partition $ A = \{ A_{i} \}_{i=0}^{n-1} \in
\mathcal{P} ( X , \mu ) $.

Consequentially the outcome of such an experiment will be a
number:
\begin{equation}
  r \in \Sigma_{n}
\end{equation}
specifying the observed atom $ A_{r} $ in our coarse-grained
observation of $ ( X, \mu ) $.

I will call such an experiment an \emph{operational observation of
$ ( X , \mu ) $ through the partition A}.

Considered another partition $ B = \{ B_{j} \}_{j=0}^{m-1} \in
\mathcal{P} ( X , \mu ) $ we have obviously that the operational
observation of $ ( X , \mu ) $ through the partition $  A \vee B $
is the conjuction of the two experiments consisting in the
operational observations of $ ( X , \mu ) $ through the
partitions, respectively, A and B.

Consequentially we may consistently call an \emph{operational
observation of $ ( X , \mu ) $ through the partition A} more
simply an \textbf{A-experiment}.

The experimental outcome of an operational observation of $ ( X ,
\mu ) $ through the partition $ A = \{ A_{i} \}_{i=0}^{n-1} \in
\mathcal{P} ( X , \mu ) $ is a classical random variable having as
distribution the stochastic vector $
\begin{pmatrix}
  \mu(A _{0} ) \\
  \vdots       \\
  \mu(A _{n-1})
\end{pmatrix} $ whose \emph{classical probabilistic information}, i.e. its
Shannon entropy,  I will call the entropy of the partition
A, according to the following:
\begin{definition}
\end{definition}
\emph{entropy of $ A = \{ A_{i} \}_{i=0}^{n-1} \in \mathcal{P} ( X
, \mu ) $}:
\begin{equation}
  H(A) := H ( \begin{pmatrix}
    \mu ( A _{0} ) \\
      \vdots       \\
    \mu ( A _{n-1} ) \
  \end{pmatrix} )
\end{equation}
It is fundamental, at this point, to observe that, given an
experiment, one has to distinguish between two conceptually
different concepts:
\begin{enumerate}
  \item  the \emph{uncertainty of the experiment}, i.e. the amount of
  uncertainty on the outcome of the experiment before to realize it
  \item  the \emph{information of the experiment}, i.e. the  amount
  of information gained by the outcome of the experiment
\end{enumerate}
As lucidly observed by Patrick Billingsley \cite{Billingsley-65},
the fact that in Classical Probabilistic Information Theory both
these concepts are quantified by the Shannon entropy of the
experiment is a consequence of the following:
\begin{theorem} \label{th:the soul of Classical Information Theory}
\end{theorem}
\emph{The soul of Classical Information Theory}:
\begin{equation}
  \text{information gained} \; = \; \text{uncertainty removed}
\end{equation}

\smallskip

Theorem \ref{th:the soul of Classical Information Theory} applies,
in particular, as to the partition-experiments we are discussing.

\medskip

Let us now consider a classical dynamical system $  CDS \, := \, (
X \, , \, \mu \, , \, T ) $.

The T-invariance of $ \mu $ implies that the partitions $ A = \{
A_{i} \}_{i=0}^{n-1} $ and $ T^{-1}A := \{ T^{-1} A_{i}
\}_{i=0}^{n-1} $ have equal probabilistic structure.
Consequentially the \emph{A-experiment} and the \emph{ $T^{-1}A
$-experiment} are replicas, \emph{not necessarily independent},
of the same experiment, made at successive times.

In the same way the \emph{$ \vee_{k=0}^{n-1} \, T^{-k} A
$-experiment} is the compound experiment consisting of n
repetitions $ A \, , \, T^{-1} A \, , \, , \cdots , \, T^{-(n-1)}A
$ of the experiment corresponding to $ A \in {\mathcal{P}}(X ,
\mu) $.

The amount of classical information per replication we obtain in
this compound experiment is clearly:
\begin{equation*}
  \frac{1}{n} \, H(\vee_{k=0}^{n-1} \, T^{-k} A )
\end{equation*}
It may be proved \cite{Kornfeld-Sinai-Vershik-00} that when n grows this
amount of classical information acquired per replication
converges, so that the following quantity:
\begin{equation}
  h( A , T ) \; := \; lim_{n \rightarrow \infty} \,  \frac{1}{n} \,
H(\vee_{k=0}^{n-1} \, T^{-k} A )
\end{equation}
exists.

In different words, we can say that $ h( A , T ) $ gives the
asymptotic rate of acquired  classical information per replication
of the A-experiment.

We can at last introduce the following fundamental notion
originally proposed  by Kolmogorov for K-systems and later
extended by Yakov Sinai to arbitrary classical dynamical systems
\cite{Kolmogorov-58}, \cite{AMS-LMS-00}, \cite{Sinai-76},
\cite{Sinai-94}, \cite{Kornfeld-Sinai-Vershik-00}:

\begin{definition} \label{def:Kolmogorov-Sinai entropy}
\end{definition}
\emph{Kolmogorov-Sinai entropy of CDS}:
\begin{equation}
  h_{CDS} \; := \; sup_{A \in {\mathcal{P}}(X , \mu)} \, h( A , T )
\end{equation}
By definition we have clearly that:
\begin{equation}
  h_{CDS} \; \geq \; 0
\end{equation}
\begin{definition} \label{def:classical chaoticity}
\end{definition}
\emph{CDS is chaotic}:
\begin{equation}
  h_{CDS} \; > \; 0
\end{equation}

The definition \ref{def:classical chaoticity} shows explicitly that the
concept of classical-chaos is an information-theoretic one:

a classical dynamical system is chaotic if there is at least one
experiment on the system that, no matter how many times we insist
on repeating it, continues to give us classical information.

That such a meaning of classical chaoticity is equivalent to the
more popular one as the sensible (i.e. exponential) dependence of
dynamics from the initial conditions  is a consequence of Pesin's
Theorem stating (under mild assumptions) the equality of the
Kolmogorov-Sinai entropy and the sum of the positive Lyapunov
exponents.

This inter-relation may be caught observing that:
\begin{itemize}
  \item if the system is chaotic we know that there is an
  experiment whose repetition definitely continues to give information: such
an information may be seen as the information on the initial condition
  that is necessary to furnish more and more with time if one want to keep
  the error on the prediction of the phase-point below a certain
  bound
  \item if the system is not chaotic the repetition of every
  experiment is useful only a finite number of times, after which
  every further repetition doesn't furnish further information
\end{itemize}

Let us now consider the issue of symbolically translating the
coarse-grained dynamics following the traditional way of
proceeding described in the second section of
\cite{Alekseev-Yakobson-1981}: given a number $ n \in {\mathbb{N}}
: n \geq 2 $ let us introduce the following:
\begin{definition}
\end{definition}
\emph{n-adic value}:

the map $ v_{n} : \Sigma_{n}^{\infty} \mapsto [ 0 , 1]$:
\begin{equation}
 v_{n} ( \bar{x} ) \; := \; \sum_{i=1}^{\infty } \frac{x_{i}}{n^{i}}
\end{equation}

\smallskip
the more usual notation:
\begin{equation}
  (0.x_{1} \cdots x_{m} \cdots
)_{n} \; := \; v_{n} ( \bar{x} ) \; \; \bar{x} \in
\Sigma_{n}^{\infty}
\end{equation}
and the following:
\begin{definition}
\end{definition}
\emph{n-adic nonterminating natural positional representation}:

the map $ r_{n} : [ 0 , 1] \mapsto \Sigma_{n}^{\infty} $:
\begin{equation}
  r_{n} ( (0.x_{1} \cdots x_{i} \cdots
)_{n} ) \;  := \; \bar{x}
\end{equation}
with the nonterminating condition requiring that the numbers of
the form $ (0. x_{1} \cdots x_{i} \overline{(n-1)})_{n} \, = \,
(0. \cdots (x_{i}+1) \bar{0})_{n}$ are mapped into the sequence $
x_{1} \cdots x_{i} \overline{(n-1)} $.

\smallskip

 Given $ n_{1} , n_{2} \in {\mathbb{N}} : \min ( n_{1} ,
n_{2} ) \geq 2 $:
\newpage
\begin{definition}
\end{definition}
\emph{change of basis from $ n_{1} $ to $ n_{2} $}:

the map $ cb_{n_{1},n_{2}} : \Sigma_{n_{1}}^{\infty} \, \mapsto \,
\Sigma_{n_{2}}^{\infty} $:
\begin{equation}
  cb_{n_{1},n_{2}} ( \bar{x} ) \; := \; r_{n_{2}}(v_{n_{1}} (
  \bar{x}))
\end{equation}
It is important to remark that \cite{Calude-02}:
\begin{theorem}
\end{theorem}
\emph{Basis-independence of randomness}:
\begin{equation}
  CHAITIN-RANDOM( \Sigma_{n_{2}}^{\infty} ) \; = \; cb_{n_{1},n_{2}} (
  CHAITIN-RANDOM(  \Sigma_{n_{1}}^{\infty} )) \; \; \forall n_{1} , n_{2} \in {\mathbb{N}} : \min ( n_{1} ,
n_{2} ) \geq 2
\end{equation}

\smallskip

 Considered a partition $ A \, = \, \{ A_{i} \}_{i = 0}^{n-1}
\in \, {\mathcal{P}}(X , \mu) $:
\begin{definition} \label{def:symbolic translator w.r.t. a partition}
\end{definition}
\emph{symbolic translator of CDS w.r.t. A}:

$ \psi_{A} \, : \, X  \rightarrow \Sigma_{n} $:
\begin{equation}
  \psi_{A} ( x  ) \; := \; i  \, : \, x \in A_{i}
\end{equation}

In this way one associates to each point of X the letter, in the
alphabet having as many letters as the number of atoms of the
considered partition, labeling the atom to which the point
belongs.

Concatenating the letters corresponding to the phase-point at
different times one can then codify  $ k \in {\mathbb{N}}$ steps
of the dynamics:
\begin{definition} \label{def:n-point symbolic translator w.r.t. a
partition}
\end{definition}
\emph{k-point symbolic translator of CDS w.r.t. A}:

$ \psi_{A}^{(k)} \, : \, X \, \rightarrow \Sigma_{n}^{k} $:
\begin{equation}
   \psi_{A}^{(k)} ( x ) \; := \; \cdot_{j = 0}^{k-1}  \psi_{A} ( T^{j} x )
\end{equation}
and whole orbits:
\begin{definition} \label{def:orbit symbolic translator w.r.t. a partition}
\end{definition}
\emph{orbit symbolic translator of CDS w.r.t. A}:

$ \psi_{A}^{(\infty)} \, : \, X \,  \rightarrow \,
\Sigma_{n}^{\infty} $:
\begin{equation}
   \psi_{A}^{( \infty)} ( x ) \; := \; \cdot_{j = 0}^{\infty}  \psi_{A} ( T^{j}
x )
\end{equation}

\medskip

The asymptotic rate of acquisition of \emph{plain Kolmogorov
complexity} of the binary sequence obtained translating
symbolically the orbit generated by $ x \in X $ through the
partition $ A \, \in \, {\mathcal{P}} ( X \, , \, \mu ) $ is
clearly given by:
\begin{definition}
\end{definition}
\begin{equation}
    B( A \, , \, x) \; := \; B (cb_{card(A),2} (\psi_{A}^{\infty}( x
    )))
\end{equation}
We saw in the  definition \ref{def:Kolmogorov-Sinai entropy} that the
\emph{Kolmogorov-Sinai entropy} was defined as $ K( A \, , \, x) $
computed on the more probabilistically-informative A-experiment;
in the same way the \emph{Brudno algorithmic entropy of x} is
defined as the value of $ B( A \, , \, x) $ computed on the more
algorithmically-informative A-experiment:
\newpage
\begin{definition} \label{def:Brudno algorithmic entropy of a point}
\end{definition}
\emph{Brudno algorithmic entropy of (the orbit starting from) x}:
\begin{equation}
  B_{CDS} (x) \; := \; sup_{A \in {\mathcal{P}}(X , \mu)} B(cb_{card(A),2}( \psi_{A}^{(\infty)}
( x )))
\end{equation}
Refering to \cite{Segre-02} for further details, let us recall
that, as it is natural for different approaches of studying a same
object, the \emph{probabilistic approach} and the
\emph{algorithmic approach} to Classical Information Theory are
deeply linked:

the partial  map $ D_{I} : \Sigma^{\star} \stackrel{ \circ } {
\mapsto } \Sigma^{\star} $ defined by:
\begin{equation}
  D_{I}( \vec{x} ) \; := \; \vec{x}^{\star}
\end{equation}
is by construction a prefix-code of pure algorithmic nature, so
that it  would be very reasonable to think that it may be optimal
only for some ad hoc probability distribution, i.e. that for a
generic probability distribution P the \emph{average code word
length of $ D_{I} $ w.r.t. P}:
\begin{equation}
  L_{ D_{I} , P } \; = \; \sum_{\vec{x} \in HALTING( D_{I} )} P(
  \vec{x}) I( \vec{x})
\end{equation}
won't achieve the optimal bound, the Shannon information H(P),
stated by the  corner stone of Classical Probabilistic
Information, i.e. the following celebrated:
\begin{theorem} \label{th:classical noiseless coding theorem}
\end{theorem}
\emph{Classical noiseless coding theorem}:
\begin{equation}
  H(P) \; \leq \; L_{P}  \; \leq \; H(P)+1
\end{equation}
(where $  L_{P} $ is the \emph{minimal average code word length
allowed by the distribution P})

Contrary,  the deep link between the \emph{probabilistic-approach}
and the \emph{algorithmic-approach} makes the miracle: under mild
assumptions about the distribution P the code $ D_{I} $ is optimal
as it is stated by the following:
\begin{theorem} \label{th:link between classical probabilistic information
and classical algorithmic information}
\end{theorem}
\emph{Link between Classical Probabilistic Information and
Classical Algorithmic Information}:

\begin{hypothesis}
\end{hypothesis}
\begin{center}
P  recursive  classical probability distribution over $
\Sigma^{\star} $
\end{center}
\begin{thesis}
\end{thesis}
\begin{equation}
  \exists c_{P} \in {\mathbb{R}}_{+} \; : \; 0 \, \leq  \,
  L_{D_{I},P} - H(P) \, \leq  \, c_{P}
\end{equation}
With an eye at the theorem \ref{th:plain-prefix insensitivity of Brudno
algorithmic entropy} it is then natural to expect that such a link
between \emph{classical probabilistic information} and
\emph{classical algorithmic information} generates a link between
the asymptotic rate of acquisition of \emph{classical
probabilistic information} and the asymptotic rate of acquisition
of \emph{classical algorithmic information} of the coarse grained
dynamics of CDS observed by repetitions of the experiments for
which each of them is maximal.

Refering to \cite{Brudno-83} for further details such a reasoning,
properly formalized, proves the following:
\begin{theorem} \label{th:Brudno theorem}
\end{theorem}
\emph{Brudno theorem}:

HP:

\begin{center}
  CDS ergodic
\end{center}

TH:

\begin{equation}
   h_{CDS} \; = \; B_{CDS} (x) \; \;  \forall - \mu - a.e. \, x
   \in X
\end{equation}

Let us now consider the \textbf{algorithmic approach to Classical
Chaos Theory} strongly supported by Joseph Ford, whose objective
is the characterization of the concept of chaoticity of a
classical dynamical system as the algorithmic-randomness of its
symbolically-translated trajectories.

To require such a condition for all the trajectories would be too
restrictive since it is reasonable to allow a chaotic dynamical
system to have a countable number of periodic orbits.

Let us then  introduce the following two notions:
\begin{definition} \label{def:strong algorithmic chaoticity}
\end{definition}
\emph{CDS is strongly algorithmically-chaotic}:
\begin{equation}
  \forall-\mu-a.e. \,  x \in X \, , \, \exists A \in {\mathcal{P}}(X ,
  \mu) \, : \, cb_{card(A),2}(\psi_{A}^{(\infty)} ( x )) \in CHAITIN-RANDOM(
  \Sigma^{\infty})
\end{equation}
\begin{definition} \label{def:weak algorithmic chaoticity}
\end{definition}
\emph{CDS is weak algorithmically-chaotic}:
\begin{equation}
  \forall-\mu-a.e. \,  x \in X \, , \, \exists A \in {\mathcal{P}}(X ,
  \mu) \, : \, cb_{card(A),2}(\psi_{A}^{(\infty)} ( x )) \in BRUDNO-RANDOM(
  \Sigma^{\infty})
\end{equation}

The difference between the  definition \ref{def:strong algorithmic
chaoticity} and the definition \ref{def:weak algorithmic chaoticity}
follows by the theorem \ref{th:Brudno randomness is weaker than Chaitin
randomness}.

Clearly the theorem \ref{th:Brudno theorem} implies the following:
\begin{corollary} \label{cor:chaoticity versus algorithmic chaoticity}
\end{corollary}
\begin{align*}
  chaoticity \; \; & = \; \; \text{weak algorithmic chaoticity} \\
  chaoticity \; \; & < \; \; \text{strong algorithmic chaoticity}
\end{align*}
that shows that the algorithmic approach to Classical Chaos Theory
is equivalent to the usual one only in weak sense.

The plethora of wrong statements found in a part of the literature
caused by the ignorance of corollary \ref{cor:chaoticity versus
algorithmic chaoticity} is anyway of little importance if
compared with the complete misunderstanding of the difference
existing among the concepts of \emph{chaoticity} and
\emph{complexity} for classical dynamical systems; with this
regards the analysis made in section \ref{sec:Introduction} may be now thoroughly formalized introducing the
following natural notions:
\begin{definition} \label{def:strong-complexity for classical dynamical
systems}
\end{definition}
\emph{CDS is strongly-complex}:
\begin{equation}
  \forall-\mu-a.e. x \in X \, , \, \exists A \in {\mathcal{P}}(X ,
  \mu) \, : \, cb_{card(A),2}(\psi_{A}^{(\infty)} ( x )) \in STRONGLY-COMPLEX(
  \Sigma^{\infty})
\end{equation}
\begin{definition} \label{def:weak-complexity for classical dynamical
systems}
\end{definition}
\emph{CDS is weakly complex}:
\begin{equation}
  \forall-\mu-a.e. x \in X \, , \, \exists A \in {\mathcal{P}}(X ,
  \mu) \, : \, cb_{card(A),2}(\psi_{A}^{(\infty)} ( x )) \in WEAKLY-COMPLEX(
  \Sigma^{\infty})
\end{equation}
One has that:
\begin{theorem} \label{th:weak-shallowness of chaotical dynamical systems}
\end{theorem}
\emph{Weak-shallowness of chaotic dynamical systems}:
\begin{center}
  CDS chaotic $ \Rightarrow $ CDS weakly-shallow
\end{center}
\begin{proof}
The thesis immediately follows combining
the theorem \ref{th:weak-shallowness of random sequences of cbits} with
the definition \ref{def:strong-complexity for classical
dynamical systems} and the definition \ref{def:weak-complexity for classical
dynamical systems}.
\end{proof}
\bigskip

  \section{The definition of the physical complexity of strings and sequences of
  qubits} \label{sec:The definition of the physical complexity of strings and sequences of
  qubits}

  The idea that the physical complexity of a quantum object has
  to be measured in terms of a quantum analogue of Bennett's
  notion of \emph{logical depth} has been first proposed by
  Michael Nielsen \cite{Nielsen-02a},\cite{Nielsen-02b}.

  Unfortunately, beside giving some general remark about the
  properties he thinks such a notion should have, Nielsen have not
  given a mathematical definition of it.

  The first step in this direction consists, in my opinion, in
  considering that, such as the notion of \emph{classical-logical-depth}
  belongs to the framework of Classical Algorithmic Information
  Theory, the notion of \emph{quantum-logical-depth} belongs to the
  framework of Quantum Algorithmic Information Theory
  \cite{Segre-02}.

  One of the most debated issues in such a discipline, first discussed by
  its father Karl Svozil
   \cite{Svozil-96} and rediscovered later  by the following
  literature \cite{Manin-99}, \cite{Vitanyi-99},
  \cite{Berthiaume-van-Dam-Laplante-01}, \cite{Gacs-01},
  \cite{Vitanyi-01}, \cite{Segre-02}, is whether the programs of
  the involved  universal quantum computers have to be strings of
  cbits or strings of qubits.

  As I have already noted in \cite{Segre-02}, anyway,  it must be observed
that, owing to the   natural bijection among the computational
basis  $ {\mathcal{E}}_{\star} $ of the \emph{Hilbert space of
qubits' strings} (notions that I am going to introduce) and $
  \Sigma^{\star}$, one can always assume that the input is a
  string of qubits while the issue, more precisely restated, is
  whether the input has (or not) to be constrained to belong to the
computational basis.

  So, denoted by $ {\mathcal{H}}_{2}  :=  {\mathbb{C}}^{2} $ the
  one-qubit's
Hilbert space (endowed with its orthonormal computational basis $
{\mathcal{E}}_{2}  := \{ | i >  ,
   i \in \Sigma \} $) , denoted by $ {\mathcal{H}}_{2}^{\bigotimes n }  :=
  \bigotimes_{k=0}^{n} {\mathcal{H}}_{2} $ the \emph{n-qubits' Hilbert
  space}, (endowed with its orthonormal computational basis $
{\mathcal{E}}_{n}  := \{ | \vec{x} >  ,
   \vec{x}  \in \Sigma^{n} \} $), denoted by $
  {\mathcal{H}}_{2}^{\bigotimes \star}  := \bigoplus_{n=0}^{\infty}
  {\mathcal{H}}_{2}^{\bigotimes n}$ the \emph{Hilbert space of qubits'
strings} (endowed with its orthonormal computational basis $
{\mathcal{E}}_{\star}  := \{ | \vec{x} >  ,
   \vec{x}  \in \Sigma^{\star} \} $) and denoted by  $
{\mathcal{H}}_{2}^{\bigotimes \infty} := \bigotimes_{n \in {\mathbb{N}}}
{\mathcal{H}}_{2} $ the \emph{Hilbert space of qubits' sequences} (endowed
with its orthonormal computational rigged-basis \footnote{as it should be
obvious,  the unusual locution \emph{rigged-basis} I am used to adopt is
simply a shortcut
   to denote that such a  "basis" has to be intended in the  mathematical
sense
it assumes when $ {\mathcal{H}}_{2}^{\bigotimes \infty} $ is
considered  as endowed with a suitable rigging, i.e. as part of a
suitable \emph{rigged Hilbert space} $ {\mathcal{S}} \subset
{\mathcal{H}}_{2}^{\bigotimes \infty} \subset {\mathcal{S}}' $ as
described in \cite{Reed-Simon-80}, \cite{Reed-Simon-75}.} $
{\mathcal{E}}_{\infty}  := \{ | \bar{x} >  ,
  \, \bar{x}  \in \Sigma^{\infty} \} $), one simply assumes that,
  instead of being a \emph{classical Chaitin universal computer},
  U is a \emph{quantum Chaitin universal computer}, i.e. a
  universal quantum computer whose input, following Svozil's original
position on the mentioned issue, is constrained to belong
  to $ {\mathcal{E}}_{\star} $ and is such that, w.r.t. the
  natural bijection among $ {\mathcal{E}}_{\star} $ and $
  \Sigma^{\star}$, satisfies the usual Chaitin constraint of
  having prefix-free halting-set.

The definition of the logical depth of a string of qubits is then
straightforward:

given a vector $ | \psi > \in {\mathcal{H}}_{2}^{\bigotimes \star}
$ and a string $ \vec{x} \in \Sigma^{\star} $:
\begin{definition} \label{def:canonical program of a vector}
\end{definition}
\emph{canonical program of $  | \psi >   $}:
\begin{equation}
  | \psi >^{\star} \; := \; \min_{<_{l} } \{
  \vec{y} \in \Sigma^{\star} \, : \, U( \vec{y} ) =  | \psi > \}
\end{equation}
\begin{definition}
\end{definition}
\emph{halting time of the computation with input  $ | \vec{x}
> $}:
\begin{equation}
    T( \vec{x} ) \; := \; \left\{%
\begin{array}{ll}
    \text{number of computational steps after which U halts on input $
\vec{x} $ }, & \hbox{if $ U(\vec{x}) = \downarrow$} \\
    + \infty, & \hbox{otherwise.} \\
\end{array}%
\right.
\end{equation}
\begin{definition} \label{def:logical depth of a string of qubits}
\end{definition}
\emph{logical depth of $ | \psi > $ at significance level s}:
\begin{equation}
    D_{s}(  | \psi >  ) \; := \; \min \{ T( \vec{y} ) \, : \, U( \vec{y}
    ) \, = \, | \psi >  \, , \, \vec{y} \text{ s-incompressible } \}
\end{equation}
\begin{definition}
\end{definition}
\emph{$ | \psi >  $ is t-deep at significance level s}:
\begin{equation}
   D_{s}(  | \psi >  ) \; > \; t
\end{equation}
\begin{definition}
\end{definition}
\emph{$ | \psi >  $ is t-shallow at significance level s}:
\begin{equation}
   D_{s}(  | \psi >  ) \; \leq  \; t
\end{equation}

I will denote the set of all the  t-deep strings of qubits  as $
t-DEEP( {\mathcal{H}}_{2}^{\bigotimes \star}  ) $.

Let us observe that a sharp distinction among \emph{depth} and
\emph{shallowness} of qubits' strings is impossible; this is
nothing but a further confirmation of the fact, so many times
shown and analyzed  in \cite{Segre-02}, that almost all the
concepts of Algorithmic Information Theory, both Classical and
Quantum, have a clear, conceptually sharp meaning only when
sequences are taken into account.

The great complication concerning sequences of qubits consists in
that their mathematically-rigorous analysis requires to give up
the simple language of Hilbert spaces passing to the more
sophisticated language of \emph{noncommutative spaces}; indeed, as
extensively analyzed in \cite{Segre-02} adopting the notion of
\emph{noncommutative cardinality} therein explicitly introduced
\footnote{following Miklos Redei's many remarks \cite{Redei-98},
\cite{Redei-01} mentioned in \cite{Segre-02},  about how Von
Neumann considered his classification of factors as a theory of
noncommutative cardinalities though he never thought, as well as
Redei, that the same $ \aleph$'s  symbolism of the commutative
case could be adopted.} , the fact that the correct noncommutative
space of qubits' sequences is the hyperfinite $ II_{1} $ factor R \footnote{
Following the notation of
\cite{Segre-02} the infinite tensor product of copies of a
generic quantum probability space $ ( A , \omega ) $ is defined as:
\begin{equation}
   \otimes_{n=1}^{\infty} ( A , \omega )  \; := \; \pi_{ \otimes_{n=1}^{\infty} \omega  } (
   \otimes_{n=1}^{\infty} A) ''
\end{equation}
where $ \pi_{\phi} $ denotes the Gelfand-Naimark-Segal
representation associated to a state $ \phi $.} :
\begin{definition}
\end{definition}
\emph{noncommutative space of qubits' sequences}:
\begin{equation}
    \Sigma_{NC}^{\infty} \; := \; \bigotimes _{n=0}^{\infty} (
    M_{2} ({\mathbb{C}} ) \, , \, \tau_{unbiased} ) \; = \; R
\end{equation}

and not the noncommutative space $ {\mathcal{B}} (
{\mathcal{H}_{2}}^{\bigotimes \infty}) $ of all the bounded linear
operators on  $ {\mathcal{H}_{2}}^{\bigotimes \infty} $ (that
could be still managed  in the usual language of Hilbert spaces)
is proved by the fact that, as it must be,  $ \Sigma_{NC}^{\infty}
$ has the \emph{continuum noncommutative-cardinality}:
\begin{equation}
     card_{NC} ( \Sigma_{NC}^{\infty} ) \; = \;  \aleph_{1}
\end{equation}
while $ {\mathcal{B}} ( {\mathcal{H}_{2}}^{\bigotimes \infty}) $
has only the \emph{countable noncommutative cardinality}:
\begin{equation}
     card_{NC} ( \Sigma_{NC}^{\star} ) \; = \;  \aleph_{0}
\end{equation}

While the definition \ref{def:strongly-deep sequence of cbits} of a
strongly-deep sequence of cbits has no natural quantum analogue,
the definition of a weakly-deep sequence of qubits is
straightforward.

Denoted by $ RANDOM( \Sigma_{NC}^{\infty} ) $ the space of all the
algorithmically random sequences of qubits, for whose
characterization I demand to \cite{Segre-02}, let us observe that
the equality between \emph{truth-table reducibility} and
\emph{recursive time bound reducibility} existing as to Classical
Computation  may be naturally imposed to Quantum Computation in
the following way:

given two arbitrary mathematical quantities x and y:
\begin{definition}
\end{definition}
\emph{x is quantum-truth-table reducible to y}:
\begin{equation}
    x \, \leq_{tt}^{Q} \, y  \; := \; \text{x is U-computable from y in
bounded U-computable time}
\end{equation}
Given a sequence of qubits $ \bar{a} \in \Sigma_{NC}^{\infty} $:
\newpage
\begin{definition} \label{def:weakly-deep sequence of qubits}
\end{definition}
\emph{$ \bar{a} $ is weakly-deep}:
\begin{equation}
    \exists \bar{b} \in RANDOM (\Sigma_{NC}^{\infty} ) \, : \neg \, ( \bar{a} \;
 \leq_{tt}^{Q} \; \bar{b} )
\end{equation}
Denoted the set of all the weakly-deep sequences of qubits as $
WEAKLY-DEEP(\Sigma_{NC}^{\infty} ) $:
\begin{definition}
\end{definition}
\emph{set of all the weakly-shallow sequences of qubits}:
\begin{equation}
  WEAKLY-SHALLOW(\Sigma_{NC}^{\infty} ) \; := \; \Sigma_{NC}^{\infty} \, -
  \, (WEAKLY-DEEP(\Sigma_{NC}^{\infty} ))
\end{equation}
It is natural, at this point, to conjecture that an analogue of the
theorem \ref{th:alternative characterization of weakly-shallow
sequences of cbits} exists in Quantum Algorithmic Information
Theory too:
\begin{conjecture} \label{con:alternative characterization of weakly-shallow
sequences of qubits}
\end{conjecture}
\emph{Alternative characterization of weakly-shallow sequences of
qubits}:
\begin{equation}
  \bar{a} \in WEAKLY-SHALLOW(\Sigma_{NC}^{\infty} ) \; \Leftrightarrow
  \; \exists \omega \in S(\Sigma_{NC}^{\infty}) \, U-computable \; : \;
\bar{a} \in \omega-RANDOM( \Sigma_{NC}^{ \infty
    })
\end{equation}
where $ \omega-RANDOM( \Sigma_{NC}^{ \infty} ) $ denotes the set
of all the $ \omega$- random sequences of qubits  w.r.t. the state
$ \omega \in S(\Sigma_{NC}^{\infty})   $ to be defined
generalizing the definition of $ RANDOM( \Sigma_{NC}^{ \infty} ) $
to states different by $ \tau_{unbiased} $  along the lines
indicated in \cite{Segre-02} as to the definition of the
\emph{laws of randomness}  $ {\mathcal{L}}^{NC}_{RANDOMNESS} (
\Sigma_{NC}^{ \infty} \, , \, \omega ) $ of the
\emph{noncommutative probability space} $  ( \Sigma_{NC}^{ \infty}
\, , \, \omega ) $.

As to sequences of qubits, the considerations made in the
section \ref{sec:Introduction} may be
thoroughly formalized,
at the prize of assuming the conjecture \ref{con:alternative characterization
of
weakly-shallow sequences of qubits} as an hypothesis, through the
following:
\begin{theorem} \label{th:weak-shallowness of random sequences of qubits}
\end{theorem}
\emph{Weak-shallowness of random sequences of qubits}:

HP:

\begin{center}
   Conjecture \ref{con:alternative characterization of weakly-shallow
sequences of qubits} holds
\end{center}

TH:

\begin{equation}
    RANDOM( \Sigma_{NC}^{ \infty} ) \, \cap \,
    WEAKLY-DEEP(\Sigma_{NC}^{\infty}) \; = \; \emptyset
\end{equation}
\begin{proof}
Since the \emph{unbiased state} $ \tau_{unbiased} $ is certainly
U-computable and by definition:
\begin{equation}
     RANDOM( \Sigma_{NC}^{ \infty} ) \; = \; \tau_{unbiased} - RANDOM(
\Sigma_{NC}^{ \infty} )
\end{equation}
the assumption of  the conjecture\ref{con:alternative
characterization of weakly-shallow sequences of qubits} as an
hypothesis  immediately leads to the thesis
\end{proof}
\newpage
   \section{The definition of the physical complexity of quantum dynamical
   systems} \label{sec:The definition of the physical complexity of quantum dynamical
   systems}

   As we have seen in section \ref{sec:The physical complexity of classical
dynamical systems} the Kolmogorov-Sinai entropy $ h_{KS}
   (CDS ) $ of a \emph{classical dynamical system} $ CDS \, := \,  ( X \, ,
\, \mu \, , \, T )
   $ has a clear physical information-theoretic meaning that we
can express in the following way:
   \begin{enumerate}
    \item an experimenter is trying  to obtain information
    about the dynamical evolution of CDS  performing repeatedly on the
system a given
    experiment $ exp \in EXPERIMENTS $,
    \item $ h(exp \, , \, CDS)  $ is  the asymptotic rate  of acquisition of
  classical information about the dynamics of CDS
    that he  acquires replicating exp
    \item $ h_{KS} (CDS ) $ is such an asymptotic rate, computed
    for the more informative possible experiment:
\begin{equation} \label{eq:physical meaning of the Kolmogorov Sinai entropy}
    h_{KS} (CDS ) \; = \; \sup_{exp \in EXPERIMENTS}  h(exp \, , \, CDS)
\end{equation}
   \end{enumerate}

Let us now pass to analyze quantum dynamical systems.

Given a  \emph{quantum dynamical system}  QDS the physical
information-theoretical way of proceeding would consist in
analyzing the same experimental situation in which an experimenter
is trying  to obtain information
    about the dynamical evolution of QDS  performing repeatedly on the
system a given
    experiment $ exp \in EXPERIMENTS $:
\begin{enumerate}
    \item to define  $ h(exp \, , \, QDS)  $ as  the asymptotic rate  of
acquisition of information about the dynamics of QDS
    that he  acquires replicating the experiment exp
    \item to define the  \emph{dynamical entropy}  of QDS  as such an
asymptotic rate, computed
    for the more informative possible experiment:
   \end{enumerate}
resulting in the following:
\begin{definition} \label{def:dynamical entropy of a quantum dynamical
system}
\end{definition}
\emph{dynamical entropy of QDS}:
\begin{equation}
    h_{d.e.} (QDS) \; = \; \sup_{exp \in EXPERIMENTS}  h(exp \, , \, QDS)
\end{equation}
\begin{definition} \label{def:quantum chaoticity}
\end{definition}
\emph{QDS is chaotic}:
\begin{equation}
     h_{d.e.} (QDS) \; > \; 0
\end{equation}

The irreducibility  of Quantum Information Theory to Classical
Information Theory, caused by the fact that the theorem \ref{th:the
soul of Classical Information Theory} doesn't extend to the
quantum case owing to the existence of some non-accessible
information about a quantum system  (as implied by the
Gr\"{o}nwald-Lindblad-Holevo Theorem) and the consequent
irreducibility of the qubit to the cbit \cite{Nielsen-Chuang-00},
\cite{Segre-02}, would then naturally lead to the physical issue
whether the information acquired by the experimenter is classical
or quantum, i.e. if $ h_{d.e.} (QDS) $ is a number of cbits or a
number of qubits.

Such a physical approach to quantum dynamical entropy  was
performed first by G\"{o}ran Lindblad  \cite{Lindblad-79} and
later refined and extended by Robert Alicki and Mark Fannes
resulting in the so called \emph{Alicki-Lindblad-Fannes entropy}
\cite{Alicki-Fannes-01}.

Many attempts to define a quantum analogue of the
\emph{Kolmogorov-Sinai entropy} pursued, instead, a different
purely mathematical approach consisting in generalizing
noncommutatively the mathematical machinery of partitions and
coarsest-refinements underlying the
definition \ref{def:Kolmogorov-Sinai entropy} obtaining
mathematical objects whose (eventual) physical meaning was
investigated subsequently.

This was certainly the case as to the
\emph{Connes-Narnhofer-Thirring entropy}, the \emph{entropy of
Sauvageot and Thouvenot} and \emph{Voiculescu's approximation
entropy} \cite{Connes-Narnhofer-Thirring-87}, \cite{Ohya-Petz-93}, \cite{Ingarden-Kossakowski-Ohya-97}, \cite{Stormer-02}.

As to the \emph{Connes-Narnhofer-Thirring entropy}, in particular,
the noncommutative analogue playing the role of the classical
partitions are the so called \emph{Abelian models} whose
(eventual) physical meaning is rather obscure since, as it has
been lucidly shown by Fabio Benatti in \cite{Benatti-93}, they  don't correspond  to physical experiments
performed on the system, since even a projective-measurement (i.e.
a measurement corresponding to a Projection Valued Measure)
cannot, in general, provide an abelian model, owing to the fact
that its reduction-formula corresponds to a decomposition of the
state of the system if and only if the measured observable belongs
to the centralizer of the state of the system.

In \cite{Segre-10} we have shown how, in the framework of Deformation Quantization, the quantum dynamical entropy may be defined very naturally simply
as the Kolmogorov-Sinai entropy of the quantum flow.

Such a definition of the quantum dynamical entropy, anway, has the defect of being applicable only to quantum dynamical systems that can be obtained
 quantizing  suitable classical hamiltonian dynamical systems.

It may be worth observing, by the way, that the non-existence of
an agreement into the scientific community as to the correct
quantum analogue of the \emph{Kolmogorov-Sinai entropy} and hence
on the definition of \emph{quantum chaoticity} shouldn't surprise,
such an agreement lacking even for the well more basic notion of
\emph{quantum ergodicity}, \emph{Zelditch's quantum ergodicity}
\cite{Zelditch-96} (more in the spirit of the original \emph{Von
Neumann's quantum ergodicity} \cite{von-Neumann-29} to which it is
not anyway clear if it reduces exactly as to quantum dynamical
systems of the form $( A \, ,  \, \omega \, , \, \alpha )$ with $
card_{NC} (A) \leq \aleph_{0} $ and $ \alpha \in INN(A) $)
differing from \emph{Thirring's quantum ergodicity}
\cite{Thirring-83} adopted both in \cite{Benatti-93} and in
\cite{Alicki-Fannes-01}.

Returning, now, to the physical approach based on the
definition\ref{def:dynamical entropy of a quantum dynamical
system}, the mentioned issue whether the dynamical entropy  $
h_{d.e.} (QDS) $ is a measure of \emph{classical information} or
of \emph{quantum information} (i.e. if it is a number of cbits or
qubits) is of particular importance as soon as one tries to extend
to the quantum domain Joseph Ford's algorithmic approach to Chaos
Theory seen in the section \ref{sec:The physical complexity of
classical dynamical systems}:
\begin{enumerate}
    \item in the former case, in fact, one should define \emph{quantum
algorithmic chaoticity} by the requirement that almost all the
trajectories, symbolically codified in a suitable way, belong to $
BRUDNO ( \Sigma^{\infty} )$ for \emph{quantum weak algorithmic
chaoticity} and  to $ CHAITIN-RANDOM( \Sigma^{\infty} ) $ for
\emph{quantum strong algorithmic chaoticity}
    \item in the latter case, instead, one should define \emph{quantum
algorithmic chaoticity} by the requirement that almost all the
trajectories, symbolically codified in a suitable way, belong to $
RANDOM ( \Sigma_{NC}^{\infty} ) $
\end{enumerate}

In any  case  one would then be tempted \cite{Segre-04b} to conjecture the
existence of a Quantum Brudno Theorem stating that the \emph{quantum chaoticity} is equivalent to \emph{quantum algorihmic chaoticity}, a task partially performed by  Fabio  Benatti, Tyll Kruger, Markus Muller, Rainer Siegmund-Schultze and Arleta Szkola   \cite{Benatti-Kruger-Muller-Siegmund-Schultze-Szkola-06},\cite{Benatti-09}.

The mentioned issue whether the dynamical entropy  $ h_{d.e.}
(QDS) $ is a measure of \emph{classical information} or of
\emph{quantum information} (i.e. if it is a number of cbits or
qubits) is of great importance also as to the definition of a
\emph{deep quantum dynamical system} (i.e. a
\emph{physically-complex quantum dynamical system}):
\begin{enumerate}
    \item in the former case, in fact, one should define a
\emph{strongly (weakly) - deep quantum dynamical system}  as a
quantum dynamical system such that  almost all its trajectories,
symbolically codified in a suitable way, belong to  $
STRONGLY-DEEP( \Sigma^{\infty})  ( WEAKLY-DEEP( \Sigma^{\infty}
))$.
    \item in the latter case, instead, one should define a
\emph{weakly-deep quantum dynamical system}  as a quantum
dynamical system such that  almost all its trajectories,
symbolically codified in a suitable way, belong to $  WEAKLY-DEEP(
\Sigma_{NC}^{\infty}   ) $
\end{enumerate}

In any case, by the Quantum Brudno Theorem and by the theorem \ref{th:weak-shallowness of random
sequences of cbits} or by the theorem \ref{th:weak-shallowness of
random sequences of qubits}, one would be almost certainly led to
a quantum analogue of theorem \ref{th:weak-shallowness of chaotical
dynamical systems} stating that a \emph{chaotic quantum dynamical
system} is  \emph{weakly-shallow}, i.e. is not \emph{physically
complex}.

In particular, given a noncommutative space (i.e a Von Neumann algebra) A, let us denote by by $ Aut_{chaotic}(A)$ the set of the chaotic automorphisms of A, by $ Aut_{complex}(A)$ the set of the deep automorphisms of A and  by $ Aut_{shallow}(A)$ the set of the shallow automorphisms of A.

Then:
\begin{theorem} \label{th:quantum chaoticity implies quantum shallowness}
\end{theorem}
\emph{Quantum chaoticity implies quantum shallowness:}
\begin{equation}
   Aut_{chaotic}(A) \, \cap \,  Aut_{complex}(A) \; = \; \emptyset
\end{equation}
\begin{proof}
\begin{itemize}
  \item Let us suppose that the quantum dynamical entropy is a measure of classical information.

  Then the thesis follows combining the Quantum Brudno Theorem with the theorem \ref{th:weak-shallowness of random
sequences of cbits}.
  \item Let us suppose, instead that the quantum dynamical entropy is a measure of quantum information.

  Then the thesis follows combining the Quantum Brudno Theorem with the theorem \ref{th:weak-shallowness of
random sequences of qubits}.
\end{itemize}
\end{proof}

\newpage
\section{Complex and shallow automorphisms of the noncommutative torus}
Let us consider the family of quantum dynamical systems of the
form $ QAT(\theta, C) \, := \,  ( T^{2}_{\theta} \, , \, \tau \,
, \, \alpha_{C} ) $ where:
\begin{equation}
    T_{\theta}^{2} \; := \; \{ \sum_{r,s \in {\mathbb{Z}}} a_{r s} u^{r} v^{s} \, : \, \{ a_{r,s} \in {\mathcal{S}} (  {\mathbb{Z}}^{2}  ) \} \}
\end{equation}
where:
\begin{equation}
     {\mathcal{S}} (  {\mathbb{Z}}^{2}  ) \; := \; \{  \{ a_{r,s} \}_{r,s \in {\mathbb{Z}} } :
     \sup_{r,s \in {\mathbb{Z}} } ( 1 + r^{2} + s^{2}) ^{k} |
     a_{ r s } |^{2} < \infty \; \; \forall k \in {\mathbb{N}} \}
\end{equation}
is the space of the rapidly decreasing double sequences and where
u and v :
\begin{equation} \label{eq:generator and relations of the torus algebra}
    v u \; = \; e^{2 \pi i \theta} u v
\end{equation}
are the generators of the \emph{torus algebra}:
\begin{equation}
  A^{2}_{\theta} \; := \; C( T^{2} ) \bowtie_{\alpha}  {\mathbb{Z}}
\end{equation}
defined as the cross-product
\cite{Gracia-Bondia-Varilly-Figueroa-01} of $ C( T^{2} ) $ and $
{\mathbb{Z}} $ w.r.t. to the following automorphism of $ C(
T^{2} ) $:
\begin{equation}
 \alpha(f) (t) \; := \; f( t + \theta )
\end{equation}

$ \tau $ is the tracial state over $ T^{2}_{\theta} $:
\begin{equation}
    \tau ( \sum_{r,s \in {\mathbb{Z}}} a_{r s} u^{r} v^{s} ) \; :=
    \; a_{00}
\end{equation}

  while $
\alpha_{C} $ is the automorphism of the noncommutative probability
space $ ( T^{2}_{\theta} \, , \tau ) $ of the form:
\begin{equation}
 u \mapsto u^{a} v^{b} \; \; v \mapsto u^{c} v^{d} \; \;  C := \left(%
\begin{array}{cc}
  a & b \\
  c & d \\
\end{array}%
\right) \in SL( 2, {\mathbb{Z}} )
\end{equation}
Since the equation \ref{eq:generator and relations of the torus algebra}
implies that:
\begin{equation}
    A_{\theta}^{2} \; \equiv \; A_{\theta + n}^{2} \; \; \forall n
    \in {\mathbb{Z}}
\end{equation}
\begin{equation}
  A_{\theta}^{2} \; \equiv \;  A_{1-\theta}^{2}
\end{equation}
the inequivalent (i.e. non $\star$-isomorphic) torus algebras are
parametrized by $ \theta \in [ 0 , \frac{1}{2} ]$.

Furthermore since:
\begin{equation}
  \theta \in {\mathbb{N}} \; \Rightarrow \;  A_{\theta}^{2} \; \equiv \; C( T^{2})
\end{equation}
\begin{equation}
    \theta \in {\mathbb{Q}} \; \Rightarrow \; {\mathcal{Z}} (
    A_{\theta}^{2} ) \; = \; C( T^{2})
\end{equation}
\begin{equation}
  \theta \notin {\mathbb{Q}} \; \Rightarrow \; {\mathcal{Z}} (
    A_{\theta}^{2} ) \; = \; \{ \lambda \mathbb{I} \, \lambda \in {\mathbb{C}}  \}
\end{equation}
we will assume from here and beyond that $ \theta \in [0,
\frac{1}{2} ] \cap ({\mathbb{R}}-{\mathbb{Q}})$ in order to deal with a factor.

\smallskip

In the following we will assume that the \emph{quantum dynamical entropy} of the automorphisms of the noncommutative torus is their \emph{Connes-Narnhofer-Thirring entropy} or their \emph{Alicki-Fannes-Lindblad entropy}:
\begin{equation}
    h( \alpha_{C} ) \; \in \; \{ h_{CNT}( \alpha_{C} ) ,  h_{AFL}( \alpha_{C} ) \} \; \; \forall C \in SL( 2 , \mathbb{Z})
\end{equation}

We remind that:
\begin{equation}
  h_{CNT} ( \alpha_{C} ) \; := \; \sup_{N \in {\mathcal{F}}(
  T^{2}_{\theta})} h_{\tau} ( \alpha_{C} , N)
\end{equation}
\begin{equation}
    h_{\tau} ( \alpha_{C} , N) \; := \; \lim_{n \rightarrow
    \infty} H_{\tau} ( N , \alpha_{C} (N) \, \cdots , \alpha_{C}^{n-1}
    (N))
\end{equation}
( $  {\mathcal{F} ( T^{2}_\theta} ) $ being the set the finite
dimensional subalgebras of $ T^{2}_{\theta} $ and where the
k-subalgebra entropy functional $ H_{\phi} ( N_{1} \, , \cdots \,
, N_{k} ) $ is defined in a very technical way about which we demand
to \cite{Benatti-93}) and, furthermore, we remind that:
\begin{equation}
    h_{AFL}  ( \alpha_{C} ) \; := \; \sup_{X \in {{\mathcal{P}} (T^{2}_{\theta \, 0} ) }
    } h ( \tau , \alpha_{C} , X )
\end{equation}
\begin{equation}
 h ( \tau , \alpha_{C} , X ) \; := \; \limsup_{n \rightarrow
 \infty} \frac{1}{n} H [ \tau ,  \alpha_{C}^{n-1}(X) \circ
 \alpha_{C}^{n-2}(X) \circ \cdots \circ  \alpha_{C} (X) \circ X ]
\end{equation}
\begin{equation}
    H[ \tau , X] \; := \; S( \rho [X])
\end{equation}
( $ {\mathcal{P}} (T^{2}_{\theta \, 0} ) $ being the set of the
operational partitions of the $ \alpha_{C} $-invariant $
\star$-subalgebra $ T^{2}_{\theta \, 0} $ of $ T^{2}_{\theta}
$, with $ S( \rho [X])$ being the Von Neumann entropy of the
\emph{correlation matrix }$ \rho [X]_{ij} := \tau ( x_{j}^{\star}
x_{i} ) $ of the operational partition of unit $ X := ( x_{1} ,
\cdots , x_{k} ) $).

\bigskip

\begin{remark}
\end{remark}
Since the notions of \emph{quantum algorithmic chaoticity} and of \emph{quantum deepness} involve a symbolical codification of the dynamics let us analyse  how such a symbolical codification is performed.

 Let us observe, first of all, that in the
Alicki-Fannes-Lindblad approach at each step an observer performs
on the system the measurement corresponding to an operational
partition of unity $ X := \{ x_{1} , \cdots , x_{k} \} $:
\begin{equation}
  \sum_{i=1}^{k} x_{i}^{\star} x_{i} \; = \; {\mathbb{I}}
\end{equation}
Such a measurement will give one among the possible results
labelled by $ \{ 1 , \cdots, k \} $.

In this way to the coarse-grained observation of the quantum
dynamics through the operational partition X it results naturally
associated a sequence $ \bar{x}' \in \Sigma_{k}^{\infty} $, i.e.
the sequence $\bar{x} \; := \; cb_{k,2} ( \bar{x}' ) \in \{ 0 ,1
\}^{\infty} $.

Let us observe, anyway, that since an operational partition is not
necessarily orthogonal so that, in general, $ x_{i}^{\star} x_{j}
\neq 0 \; i \neq j $, the information obtained by the measurement
associated to the operational partition $ X = ( x_{1} , \cdots ,
x_{k} ) $ is  higher than the Shannon entropy:
\begin{equation}
    S( \vec{P} ) \; = \; - \sum_{i=1}^{k} p_{i} \log_{2} p_{i}
\end{equation}
where $ \vec{P} = ( p_{1} , \cdots , p_{k} ) $ is the probability
vector for the possible outcomes.

Let us observe that the measurement associated to X corresponds to
a coarse-grained observation of the underling quantum probability
space $ (T^{2}_{\theta} , \tau)  $ as the $ k^{2}$ dimensional
quantum probability space $ ( M_{k}({\mathbb{C}}) , \tau[X])$
where $ \tau[X] $ is the normal state over $ M_{k}({\mathbb{C}}) $
with density matrix $ \rho [X]_{ij} $, as induced by the
coarse-graining completely positive  unity preserving map $
\Gamma_{X} : M_{k} ({\mathbb{C}}) \mapsto T^{2}_{\theta}$:
\begin{equation}
    \Gamma_{X} ( [ a_{ij} ] ) \; := \; \sum_{i,j=1}^{k} a_{ij}
    x_{i}^{\star} x_{j}
\end{equation}

Let us now consider the coarse-grained observation of the
dynamics.

We have, at this purpose, to recall some notion about operator
algebras.

Given an $ n \in {\mathbb{N}} $ let us denote by $ \Sigma_{NC \, n
} := M_{n} ( {\mathbb{C}}  )$ the noncommutative alphabet of n
letters and let us introduce the following:
\newpage
\begin{definition}
\end{definition}
\emph{sequences over  $ \Sigma_{NC \, n } $}:
\begin{equation}
     \Sigma_{NC \, n }^{\infty} \; := \; \otimes_{n=1}^{\infty} ( \Sigma_{NC \, n
     } , \tau_{n} )
\end{equation}
where $ \tau_{n} $ is the tracial state over $  \Sigma_{NC \, n
     } $.

Let us now recall that \cite{Kadison-Ringrose-97a},
\cite{Kadison-Ringrose-97b}:
\begin{theorem} \label{th:the spaces of sequences over a finite noncommutative alphabet are all the same}
\end{theorem}
\begin{equation}
 \Sigma_{NC \, n_{1} }^{\infty} \; = \; \Sigma_{NC \, n_{2}
 }^{\infty} \; = \; R \; \; \forall n_{1} , n_{2} \in
 {\mathbb{N}}_{+}
\end{equation}
where we have identified $\star$-isomorphic $ W^{\star}$-algebras.

Let us now observe that the orbit associated to an initial
condition $ a \in M_{k} ({\mathbb{C}}) $ will be a  $ \bar{a} \in
\otimes_{n=1}^{\infty}  ( M_{k}({\mathbb{C}}) , \tau[X])$.

We have clearly that:
\begin{equation}
   \tau[X] = \tau_{k} \; \Rightarrow \; \otimes_{n=1}^{\infty}  ( M_{k}({\mathbb{C}}) , \tau[X]) \; =
    \; \Sigma_{NC}^{\infty} 
\end{equation}
where $ \tau_{k} $ denotes the tracial state over $
M_{k}({\mathbb{C}}) $ and where, following the notation of
\cite{Segre-02}, \cite{Segre-04b}, $ \Sigma_{NC} \; := \;
\Sigma_{NC,2} $ is the one qubit alphabet, i.e. the binary
noncommutative alphabet.

\bigskip

An explicit computation of such dynamical entropies shows that \cite{Benatti-93}, \cite{Alicki-Fannes-01}, \cite{Benatti-09}:
\begin{theorem} \label{th:theorem on the dynamical entropies of the noncommutative torus}
\end{theorem}
\begin{equation}
    h_{KS} ( C) \; = \; h_{CNT} (\alpha_{C}) \; = \; h_{AFL} (\alpha_{C}) \; = \; \left\{
                                                                                    \begin{array}{ll}
                                                                                      \log \lambda, & \hbox{if $  Tr ( C)  > 2 $;} \\
                                                                                      0, & \hbox{ if $  Tr ( C)  \leq 2 $.}
                                                                                    \end{array}
                                                                                  \right.
\end{equation}
where $ \lambda $ is the greatest eigenvalue of the matrix C.

Let us now assume that the \emph{quantum dynamical entropy} of an automorphism of the noncommutative torus is its \emph{Connes-Narnhofer-Thirring entropy} or its \emph{Alicki-Fannes-Lindblad entropy}:
\begin{equation}
    h_{d.e}( \alpha_{C} ) \; \in \; \{ h_{CNT}( \alpha_{C}  ) ,  h_{AFL}( \alpha_{C}  ) \} \; \; \forall C \in SL( 2 , \mathbb{Z} )
\end{equation}
Then obviously:
\begin{corollary}
\end{corollary}
\begin{equation}
  Aut_{chaotic}( T_{\theta}^{2} ) \; = \; \{ \alpha_{C} \, : \,  Tr ( C)  > 2 \}
\end{equation}

Furthermore:
\begin{theorem} \label{th:some shallow automorphisms}
\end{theorem}
\begin{equation}
    \{ \alpha_{C} \, : \,  Tr ( C)  > 2 \} \, \cap \, Aut_{complex}( T_{\theta}^{2}) \; = \; \emptyset
\end{equation}
\begin{proof}
Then thesis immediately follows combining the theorem \ref{th:theorem on the dynamical entropies of the noncommutative torus} with the theorem \ref{th:quantum chaoticity implies quantum shallowness}.
\end{proof}

\newpage
\begin{example}
\end{example}
\emph{The classical and quantum Arnold cats are not complex:}

Let us consider in particular the automorphism of the torus $ T^{2} $ identified by the matrix:
\begin{equation}
    C_{cat} \; := \; \left(
                       \begin{array}{cc}
                         1 & 1 \\
                         1 & 2 \\
                       \end{array}
                     \right)
\end{equation}
usually called in the literature the \emph{Arnold cat map}.

Since $  Tr(C_{cat})  \, = \, 3 > 2 $ and $ \lambda \, = \, \frac{3+ \sqrt{5}}{2} $ it follows that:
\begin{equation}
    h_{KS}( C_{cat}) \; = \; h_{CNT}( \alpha_{C_{cat}}) \; = \; h_{AFL}( \alpha_{C_{cat}}) \; = \; \log ( \frac{3+ \sqrt{5}}{2})
\end{equation}

By the theorem \ref{th:some shallow automorphisms} it follows not only that such a classical dynamical system is not complex but also that the quantum dynamical system
 $ \alpha_{C_{cat}}$, usually called the \emph{quantum Arnold cat map} is not complex too.

\bigskip

\newpage
\section{Notation}

\bigskip
\begin{tabular}{|c|c|}
  $ \forall $ & for every (universal quantificator) \\
  $ \forall-\mu-a.e. $ & for $ \mu-almost$ every \\
  $ \exists $ & exists (existential quantificator) \\
  $ \exists \; ! $ & exists and is unique \\
  $ x \; = \; y $ & x is equal to y \\
  $ x \; := \; y $ & x is defined as y \\
  $ x \; \equiv \; y $ & x is equivalent to y \\
  $ \neg p $ & negation of p \\
  $ \Sigma $ & binary alphabet $ \{ 0 , 1 \} $  \\
  $ \Sigma_{n} $ & n-letters' alphabet \\
  $ \Sigma_{n}^{\star} $ & strings on the alphabet $ \Sigma_{n} $  \\
  $ \Sigma_{n}^{\infty} $ &  sequences on the alphabet $ \Sigma_{n} $ \\
  $ \vec{x} $ & string \\
  $ | \vec{x} | $ &  length of the string $ \vec{x} $ \\
  $ <_{l} $ & lexicographical ordering on  $ \Sigma^{\star} $ \\
  $ string(n) $ & $ n^{th} $ string in lexicographic order \\
  $ \bar{x} $ & sequence \\
  $ \cdot $ & concatenation operator \\
  $ x_{n} $ & $ n^{th} $ digit of the string $\vec{x} $ or of the sequence $
\bar{x} $ \\
  $ \vec{x}(n) $ & prefix of length n of the string $ \vec{x} $ or of the
sequence $ \bar{x} $  \\
  $ \vec{x}^{\star} $ & canonical string of the string  $ \vec{x}$
  \\
$ K( \vec{x} ) $ & plain Kolmogorov complexity of the string $
\vec{x} $ \\
$ I( \vec{x} ) $ & algorithmic information of the string $
\vec{x} $ \\
$ U( \vec{x} ) \downarrow $ & U halts on input $ \vec{x} $ \\
$ D_{s} ( \vec{x} ) $ & logical-depth of $ \vec{x} $ at
significance level s \\
$ t-DEEP( \Sigma^{\star} ) $ & t-deep strings of cbits \\
$ t-SHALLOW  ( \Sigma^{\star} ) $ & t-shallow strings of cbits \\
$ REC-MAP ( {\mathbb{N}} \, , \, {\mathbb{N}} ) $ &  (total)
recursive functions over $  {\mathbb{N}} $ \\
$ \leq_{T} $ & Turing reducibility \\
$ \leq_{T}^{P} $ & polynomial time Turing reducibility \\
$ \leq_{tt} $ & truth-table reducibility \\
$ CHAITIN-RANDOM (\Sigma^{\infty} ) $ & Martin L\"{o}f - Solovay
- Chaitin random sequences of cbits \\
$ HALTING( \mu ) $ & halting set of the measure $ \mu $ \\  \hline
\end{tabular}

\newpage
\begin{footnotesize}

\begin{tabular}{|c|c|}
$ \mu_{Lebesgue} $ & Lebesgue measure \\
$ \mu-RANDOM( \Sigma^{\infty} ) $ &  Martin L\"{o}f random
sequences of cbits w.r.t. $ \mu $ \\
$ B( \bar{x} ) $ & Brudno algorithmic entropy of the sequence $
\bar{x} $ \\
$ BRUDNO (\Sigma^{\infty} ) $ & Brudno random sequences of cbits \\
$ STRONGLY-DEEP (\Sigma^{\infty} ) $ & strongly-deep sequences of
cbits \\
$ STRONGLY-SHALLOW (\Sigma^{\infty} ) $ & strongly-shallow
sequences of cbits \\
$ WEAKLY-DEEP (\Sigma^{\infty} ) $ & weakly-deep sequences of
cbits \\
$ WEAKLY-SHALLOW (\Sigma^{\infty} ) $ & weakly-shallow sequences
of cbits \\
$ {\mathcal{P}} ( X \, , \, \mu ) $ & (finite, measurable)
partitions of $ ( X \, , \, \mu ) $ \\
$ \preceq $ & coarse-graining relation on partitions \\
$ A \bigvee B $ & coarsest refinement of A and B \\
$ h_{CDS} $ & Kolmogorov-Sinai entropy of CDS \\
$ \psi_{A} $ & symbolic translator w.r.t. A \\
$ \psi_{A}^{(k)} $ & k-point symbolic translator w.r.t. A \\
$ \psi_{A}^{(\infty)} $ & orbit symbolic translator w.r.t. A \\
$ cb_{n_{1},n_{2}} $ & change of basis from $ n_{1} $ to $ n_{2} $  \\
$ B_{CDS} (x) $ & Brudno algorithmic entropy of x's  orbit in CDS \\
$ H(P) $ & Shannon entropy of the distribution P \\
$ L_{D,P} $ & average code-word length w.r.t. the code D and the
distribution P \\
$ L_{P} $ & minimal average code-word length w.r.t. the
distribution P \\
$ {\mathcal{H}}_{2} $  & one-qubit's Hilbert space \\
$ {\mathcal{H}}_{2}^{\bigotimes n} $  & n-qubits' Hilbert space \\
$  {\mathcal{E}}_{n} $ & computational basis of $
{\mathcal{H}}_{2}^{\bigotimes n} $ \\
$ {\mathcal{H}}_{2}^{\bigotimes \star} $  & Hilbert space of
qubits' strings \\
$  {\mathcal{E}}_{\star} $ &  computational basis of $
{\mathcal{H}}_{2}^{\bigotimes \star} $ \\
$  {\mathcal{H}}_{2}^{\bigotimes \infty } $  & Hilbert space of
qubits' sequences \\
$  {\mathcal{E}}_{\infty} $ &  computational rigged-basis of $
{\mathcal{H}}_{2}^{\bigotimes \infty} $ \\
$ {\mathcal{B}} ( {\mathcal{H}} ) $ & bounded linear operators on
$ {\mathcal{H}} $ \\
$ | \psi >^{\star} $ & canonical program of $ | \psi > $ \\
$ D_{s} ( | \psi > ) $  & logical depth of $ | \psi > $  at
significance level s \\
$ t-DEEP( {\mathcal{H}}_{2}^{\bigotimes \star}  ) $ & t-deep
strings of qubits \\
$ t-SHALLOW( {\mathcal{H}}_{2}^{\bigotimes \star}  ) $ & t-shallow
strings of qubits \\
$ S(A) $ & states over the noncommutative space A \\
$ \mathcal{Z}(A) $ & center of the noncommutative space A \\
$ AUT(A) $ & automorphisms of the noncommutative space A \\
$ AUT_{chaotic}(A) $ & chaotic automorphisms of the noncommutative space A \\
$ AUT_{complex}(A) $ & complex automorphisms of the noncommutative space A \\
$ AUT_{shallow}(A) $ & shallow automorphisms of the noncommutative space A \\
$ INN(A) $ & inner automorphisms of the noncommutative space A \\
$ card_{NC} (A) $ & noncommutative cardinality of the noncommutative space  A \\
$ \tau_{unbiased} $ & unbiased noncommutative probability
distribution \\
$ \Sigma_{NC}^{\infty} $ &  noncommutative space of qubits' sequences \\
$ \Sigma_{NC \, n }^{\infty} $ & sequences over the noncommutative alphabet with n letters \\
R & hyperfinite $ II_{1} $ factor \\
$ RANDOM ( \Sigma_{NC}^{\infty} ) $ &  random sequences of qubits
\\
$ \leq_{tt}^{Q} $ & quantum truth-table reducibility \\
$ WEAKLY-DEEP (\Sigma_{NC}^{\infty} ) $ & weakly-deep sequences of
qubits \\
$ WEAKLY-SHALLOW (\Sigma_{NC}^{\infty} ) $ & weakly-shallow
sequences
of qubits \\
$ {\mathcal{L}}_{RANDOMNESS}^{NC} ( A , \omega ) $ & laws of
randomness of $ ( A \, , \, \omega ) $ \\
$ \omega - RANDOM ( \Sigma_{NC}^{\infty} ) $ &  random sequences
of qubits w.r.t. $ \omega $ \\
$ h_{d.e.} (QDS) $ & dynamical entropy of the quantum dynamical
system QDS \\
$ h_{CNT} (QDS) $ & Connes-Narnhofer-Thirring entropy of the quantum dynamical
system QDS \\
$ h_{AFL} (QDS) $ & Alicki-Fannes-Lindlad entropy of the quantum dynamical
system QDS \\
$ A \bowtie_{\alpha} G $ & cross-product of A and G induced by $ \alpha $ \\
$ A_{\theta}^{2} $  & torus algebra \\
$  T_{\theta}^{2} $  & noncommutative torus \\
$ C_{Cat} $ & matrix of the Arnold cat map \\
i.e. & id est \\
e.g. & exempli gratia \\ \hline
\end{tabular}
\end{footnotesize}

\end{document}